\documentclass[twocolumn,showpacs,amsmath,nofootinbib,pra,aps,amssymb]{revtex4-1}

\usepackage[T1]{fontenc}
\usepackage[utf8]{inputenc}
\usepackage{times}
\usepackage{color}
\usepackage{physics} 
\usepackage{array}
\usepackage{amssymb,amsmath}
\usepackage{amsbsy}
\usepackage[pdftex]{graphicx}
\usepackage{bm}
\usepackage{float}
\usepackage{dcolumn} 
  \usepackage{dsfont}
\usepackage[unicode,breaklinks]{hyperref}
\hypersetup{
    unicode=true,
    plainpages=false, 
    colorlinks=true,
    linkcolor=blue,
    citecolor=blue,
    filecolor=black,
    urlcolor=blue
}
\usepackage{url}
\usepackage{verbatim}

\begin{document}

\title{Properties of a nematic spin vortex in an antiferromagnetic spin-1 Bose-Einstein condensate}
	\author{Andrew P.~C.~Underwood}	
	\affiliation{ Dodd-Walls Centre for Photonic and Quantum Technologies, New Zealand}
	\affiliation{ Department of Physics, University of Otago, Dunedin 9016, New Zealand} 
	\author{D.~Baillie}
		\affiliation{ Dodd-Walls Centre for Photonic and Quantum Technologies, New Zealand}
	\affiliation{ Department of Physics, University of Otago, Dunedin 9016, New Zealand} 
	\author{P.~Blair~Blakie}
		\affiliation{ Dodd-Walls Centre for Photonic and Quantum Technologies, New Zealand}
	\affiliation{ Department of Physics, University of Otago, Dunedin 9016, New Zealand} 
	\author{H.~Takeuchi}
	\affiliation{ Department of Physics and Nambu Yoichiro Institute of Theoretical and Experimental Physics (NITEP), Osaka City University, Osaka 558-8585, Japan}

 \date{\today}
\begin{abstract}  
A spin-1 condensate with antiferromagnetic interactions supports nematic spin vortices in the easy-plane polar phase. These vortices have a $2\pi$ winding of the nematic director, with a core structure that depends on the quadratic Zeeman energy. 
 We characterize the properties of the nematic spin vortex in a uniform quasi-two-dimensional system. We also obtain the vortex excitation spectrum and use it to quantify its stability against dissociating into two half-quantum vortices, finding a parameter regime where the nematic spin vortex is dynamically stable. These results are supported by full dynamical simulations.
\end{abstract}
\maketitle

\section{Introduction}
Spinor Bose-Einstein condensates are superfluid quantum gases with spin degrees of freedom. These can exist in various spin-ordered phases depending on the nature of the inter-particle interactions and quadratic Zeeman shift \cite{Ho1998a,Ohmi1998a,Stamper-Kurn1998,Kawaguchi2012R,StamperKurn2013a}. Quantized vortices are regarded a hallmark of superfluidity and often play a unifying role in nonequilibrium dynamics \cite{Bray1994,Sadler2006a,Kudo2015a,Williamson2016a}. The rich order parameter symmetries of spinor condensates give rise to an array of different spin vortices. Emerging experimental capabilities to produce and monitor the dynamics of spin vortices \cite{Seo2015a,Seo2016a,Chen2018a} motivate the need for a better understanding of their properties and interactions (e.g.~see \cite{Turner2009,Eto2011a,Williamson2016c,Kasamatsu2016a}). 

In this paper we consider a spin-1 antiferromagnetic condensate with polar (nematic spin) order which can be characterized by a director, 
i.e.~a preferred unoriented axis in spin-space \cite{Zibold2016a,Symes2017b}.
 When the quadratic Zeeman energy is negative, the condensate is in the easy-plane polar (EPP) phase where the director lies in the plane transverse to the magnetic field.  In this case, the director breaks the continuous rotational symmetry and the order-parameter manifold supports various spin vortices as topological defects (e.g.~see \cite{Isoshima2002a,Lovegrove2016a,Lovegrove2016a}). A significant amount of attention has been given to half-quantum vortices (HQVs) \cite{Leonhardt2000a}, which consist of mass and nematic spin current circulation and have recently been prepared in experiments \cite{Seo2015a,Seo2016a}. The role of HQVs in post-quench dynamics \cite{Kang2017a,Symes2017b,Kang2019a,Symes2018a} and the interactions between HQVs \cite{Eto2011a,Kasamatsu2016a,Seo2016a} have been studied.
 Here we focus on a second type of vortex in the EPP phase: a pure spin-vortex (i.e.~without mass current), which we refer to as the nematic spin vortex (NSV) (also see \cite{Lovegrove2014a,Lovegrove2016a,Choi2012a,Choi2012b}). These NSVs can decay by splitting into a pair of HQVs, and thus an important consideration is their stability. We note previous work has considered NSVs in a harmonically trapped system, and explored their energetic stability under external rotation and varying magnetization  \cite{Lovegrove2014a,Lovegrove2016a}.

In this paper we develop theory for a NSV in an infinite uniform quasi-two-dimensional (quasi-2D) EPP condensate. This allows us to describe the core structure and excitation spectrum using two parameters: the quadratic Zeeman energy scaled by the chemical potential, and the ratio of the spin-dependent to spin-independent interactions. We determine a critical value $q_c$ of the quadratic Zeeman energy where the NSV undergoes a continuous transition from having a normal (unfilled) core to having a core filled by an easy-axis polar (EAP) component. We quantify the dissociation instability of the NSV by solving the Bogoliubov-de Gennes (BdG) equations, and by performing dynamical simulations. Importantly we find that at small negative values of the quadratic Zeeman energy, the NSV is dynamically stable.   

The structure of the paper is as follows. In Sec.~\ref{Sec:BKGTHRY} we introduce the background theory for the spin-1 system and overview the vortices in the EPP phase. In Sec.~\ref{Sec:NVS} we specialize the theory to NSV stationary states in a quasi-2D system, and present numerical  results for the vortex properties.
In Sec.~\ref{Sec:Excitations} we formulate the BdG equations for the NSV and present a phase diagram characterizing the strength of dynamic instabilities.  Using dynamical simulations of the spin-1 system in a flat-bottomed trap we verify the splitting instability emerging from the dynamic instability in Sec.~\ref{Sec:dynamics}. We conclude in Sec.~\ref{Sec:conclude}.

 \section{Background theory}\label{Sec:BKGTHRY}
\subsection{General formalism for spin-1 BECs}\label{Sec:GenFormalism}
A spin-1 condensate is described by the spinor field
\begin{align}
\bm{\Psi}\equiv [\Psi_1,\Psi_0,\Psi_{-1}]^\mathrm{T},\label{Eqpsisph}
\end{align} 
with the three components representing the condensate amplitude in the spin levels $m=1,0,-1$, respectively, where $m$ is the quantum number associated with the $z$-component of spin.
In weak fields the short-ranged contact interactions between atoms are rotationally invariant with a Hamiltonian density  
\begin{align}\label{spinH}
\mathcal{H}_{\mathrm{int}}= \frac{c_0}{2}n^2+\frac{c_1}{2}|\vec{F}|^2. 
\end{align}
Here the first term, with coupling constant $c_0$, describes the density dependent interactions, where $n\equiv\bm{\Psi}^\dagger\bm{\Psi}$ is the total density. The second term describes the spin-dependent interactions, where $c_1$ is the spin-dependent coupling constant, $\vec{F}\equiv \bm{\Psi}^\dagger\vec{f}\bm{\Psi}$ is the spin density, and  $\vec{f} \equiv (\check{f}_x,\check{f}_y,\check{f}_z)$  are the  spin-1 matrices.  In addition, we consider the presence of a (uniform) quadratic Zeeman shift. Taking the field to be along $z$ this is described by 
\begin{align}
\mathcal{H}_{\mathrm{QZ}}=q\bm{\Psi}^\dagger \check{f}_z^2\bm{\Psi}=q\left(|\Psi_{1}|^{2}+|\Psi_{-1}|^{2}\right).\label{HQZ}
\end{align} 
In practice the coefficient $q$ is readily changed in experiments using microwave dressing (e.g.~see \cite{Gerbier2006a,Leslie2009a}).
We note that the uniform linear Zeeman term can be removed using a gauge rotation, and can be neglected. The spin properties also depend on the  (conserved) $z$-magnetization $M_{z}=\int\dd{V} F_{z}$ of the system. Here we consider only $M_{z}=0$.

The case of antiferromagnetic interactions, where $c_1>0$, is realized with $^{23}$Na atoms in their lowest hyperfine manifold.
 Here the condensate prefers to minimize the spin-density to reduce the spin-dependent interaction energy. For a condensate of uniform density $n_{b}$ and spin-density $\vec{F}=\vec{0}$ the spinor is in a polar state 
\begin{align}
\bm{\Psi}_\text{P}=\left[\begin{array}{c}\Psi_{1}\\ \Psi_0 \\ \Psi_{-1}\end{array}\right]
=\sqrt{n_{b}}e^{i\theta}\left[\begin{array}{c} \frac{-d_x+id_y}{\sqrt{2}} \\
d_z \\ \frac{d_x+id_y}{\sqrt{2}} \end{array}\right],\label{psiPolar}
\end{align}
where the real unit vector $\vec{d}=(d_x,d_y,d_z)$ is the nematic director and $\theta$ is the global phase. 
 Noting that $\bm{\Psi}_\text{P}$ is invariant under $\theta\to\theta+\pi$ and $\vec{d}\to-\vec{d}$, we see that $\vec{d}$ defines a preferred axis in spin space, but not a preferred direction along that axis. The ground state orientation of $\vec{d}$ is determined by the quadratic Zeeman energy, which is given by $\mathcal{H}_{\mathrm{QZ}}=qn_{b}(1-d_z^2)$. Thus for $q>0$ the system maximizes $d_z^2$, by being in the EAP phase, i.e.~$\bm{\Psi}_\text{EAP}=\sqrt{n_{b}}e^{i\theta}(0,1,0)^T$. The case of interest in this paper is the EPP phase for $q<0$ where $\vec{d}=(d_x,d_y,0)$. In addition to the global phase, the EPP ground state also breaks a U(1) symmetry in spin space. This can be seen from the director, which can be written as $\vec{d}= (\cos\varphi,\sin\varphi,0)$, i.e.~$\bm{\Psi}_\text{EPP}=\sqrt{\frac{n_{b}}{2}}e^{i\theta}(-e^{-i\varphi},0,e^{i\varphi})^T$, where $\varphi$ is the angle the director takes with respect to the $x$-axis. 
 
Note that in the EPP phase we have $\Psi_0=0$ and the system is effectively a two-component condensate. Indeed, several studies of the relevant EPP vortices we present in the next subsection have been performed in a two-component (or binary) condensate. For completeness we briefly mention the mapping of the spinor parameters onto an equivalent binary system. To do this we note that interaction Hamiltonian density $\mathcal{H}_{\mathrm{int}}$ may be expressed in the binary form
\begin{align}
\mathcal{H}_{\mathrm{int}}=\frac{1}{2}\sum_{i,j=-1,1}g_{ij}|\Psi_{i}|^{2}|\Psi_{j}|^{2}, 
\end{align}
with intra-species coupling constant $g_{ii}=c_{0}+c_{1}$ (identical for both components) and interspecies coupling constant $g_{1,-1}=c_{0}-c_{1}$. For the antiferromagnetic case $g_{1,-1}<g_{ii}$ and the components are miscible \cite{Mineev1974a}.

 \subsection{EPP phase vortex classification by winding numbers}
 Here we consider a quasi-2D spinor gas where the order parameter manifold of the EPP phase permits vortices as point defects.
To give the basic structure of such vortex states we write the wave function on the $xy$-plane with the vortex core taken at the origin.  Sufficiently far from the core the general vortex state is of the form 
 \begin{align}
\bm{\Psi}_\text{V}=\sqrt{\frac{n_{b}}{2}}e^{i\sigma_\text{M}\phi}\left[\begin{array}{c}-e^{-i\sigma_\text{S}\phi}\\  0 \\ e^{i\sigma_\text{S}\phi}\end{array}\right]\label{psiv}
\end{align}
where  $\phi$  is the azimuthal angle in the $xy$-plane [i.e.~from $\bm{\Psi}_\text{EPP}$, taking $\theta$ and $\varphi$ to vary as $\sigma_\text{M}\phi$ and  $\sigma_\text{S}\phi$ in space, respectively]. Here $\sigma_\text{M}$ and $\sigma_\text{S}$  are the winding numbers associated with the mass and spin current around the vortex, respectively (e.g.~see \cite{Kudo2015a}). Combining the circulations for the $m=\pm1$ components we see that $\sigma_\text{M}\pm\sigma_\text{S}$ must both be integers for the field to be singled valued. There are 8 non-trivial cases with $|\sigma_\text{M}\pm\sigma_\text{S}|\le1$ which define the elementary vortices of interest.

\subsubsection{HQV}
The HQVs come in 4 types with $(\sigma_\text{M},\sigma_\text{S})=(\pm\frac{1}{2},\pm\frac{1}{2})$ and $(\pm\frac{1}{2},\mp\frac{1}{2})$, thus exhibiting both spin and mass currents. In these vortices the director completes a $\pi$ winding around the vortex core. These vortices have recently been studied in experiments and observed to be long lived  topological defects \cite{Seo2015a,Seo2016a}. Indeed, HQVs are expected to be stable topological defects of the EPP phase and have been the focus of studies of nonequilibrium dynamics in this phase (e.g.~see \cite{Kang2017a,Symes2017b,Symes2018a}).  Experiments have also observed annihilation events between suitable pairs of HQVs \cite{Seo2016a}, e.g.~the pair $\{(\frac{1}{2},\frac{1}{2}),(-\frac{1}{2},-\frac{1}{2})\}$ or $\{(\frac{1}{2},-\frac{1}{2}),(-\frac{1}{2},\frac{1}{2})\}$  can mutually annihilate.

\subsubsection{Mass vortex}
The mass vortex has $(\sigma_\text{M},\sigma_\text{S})=(\pm1,0)$, with mass current but no spin current.  These vortices have been prepared in the EPP phase \cite{Seo2015a}, but were observed to rapidly decay by dissociation into two HQVs, i.e.~
\begin{align}
(\pm1,0)\to (\pm\tfrac{1}{2},\pm\tfrac{1}{2}) + (\pm\tfrac{1}{2},\mp\tfrac{1}{2}),
\end{align}
as anticipated by theoretical studies \cite{Ji2008a,Eto2011a,Lovegrove2012a}.

\begin{figure}[htbp] 
   \centering
   \includegraphics[width=\linewidth]{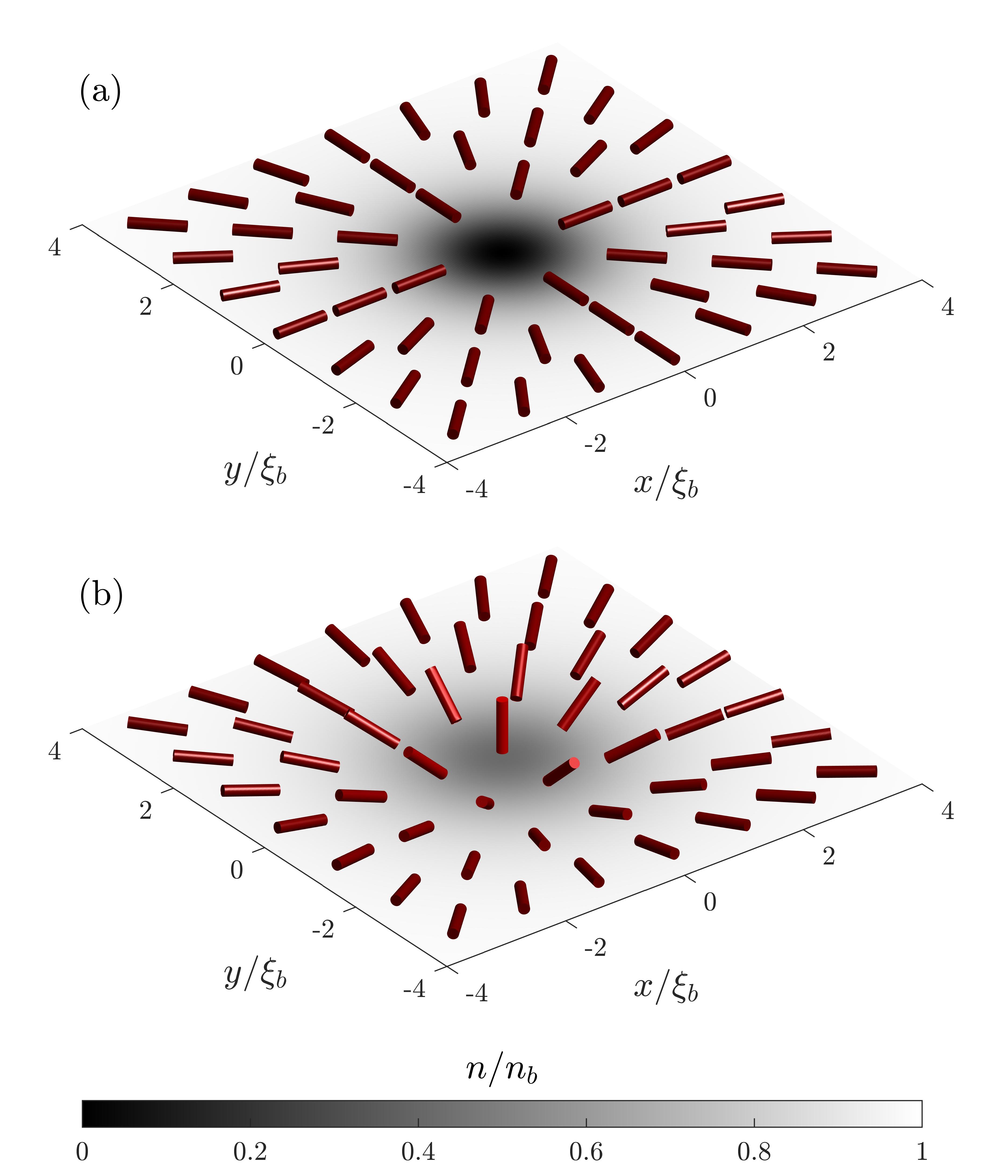}%{NematicTexture_pc_2.png}  
   \caption{The basic structure of the NSV at the origin in a quasi-2D EPP phase condensate.  (a) Normal-core NSV and (b) a polar-core NSV. Cylindrical rods indicate the orientation of the nematic director $\vec{d}$. Far from the vortex core the director lies in the plane and undergoes a $2\pi$ rotation as we complete a closed loop around the core. The background shaded colours indicate the total density. }
   \label{fig:schematic}
\end{figure}

\subsubsection{NSV}\label{Sec:IntroNSV}
The NSV has $(\sigma_\text{M},\sigma_\text{S})=(0,\pm1)$,  thus exhibits a spin current but no mass current. 
Similar to the mass vortex the NSV can potentially dissociate into two HQVs \cite{Ishino2013a} as 
\begin{align}
(0,\pm1)\to(\pm\tfrac{1}{2},\pm\tfrac{1}{2})+(\mp\tfrac{1}{2},\pm\tfrac{1}{2}).
\end{align}
We consider the stability to this dissociation process in Sec.~\ref{Sec:Excitations}.

\section{Stationary NSV Solutions}\label{Sec:NVS}

Here, we investigate the structure of the NSV  core. Examples of the two types of NSV are shown in Fig.~\ref{fig:schematic}. We see that completing a loop around the core (at a radius sufficiently far from the core) the director remains in the easy-plane manifold and completes a $2\pi$ winding. The two examples differ in their structure near the vortex core. The case in Fig.~\ref{fig:schematic}(a), which we refer to as the \textit{normal-core} NSV, has the total density vanish at the vortex core, with the director always remaining in the plane (i.e.~$d_z=0$). The case in Fig.~\ref{fig:schematic}(b), which we refer to as the \textit{polar-core} NSV, instead has a finite density at the vortex core. Here the director moves out of the EPP order parameter manifold (i.e.~tilts out of the plane near the core) showing the emergence of the EAP state within the core. We adopt the name polar core for this case, with polar being a conventional name for the EAP phase. 
Here we show that there is a continuous transition between these two types of NSV as $q$ changes. 

   \subsection{Spin-1 Gross-Pitaevskii equation}
The evolution of a spin-1 condensate is described by the Gross-Pitaevskii equation (GPE) \begin{align}
i\hbar\frac{\partial {\bm{\Psi}}}{\partial t}=\check{\mathcal{L}}\bm{\Psi},\label{tdGPE}
\end{align}
where the $\check{\mathcal{L}}$ is the nonlinear GPE operator
\begin{align}
\check{\mathcal{L}}=-\frac{\hbar^2\nabla^2}{2M}\mathds{1} +q\check{f}_z^2+c_0n\mathds{1} +c_1\sum_{\alpha}F_{\alpha}\check{f}_{\alpha},
\end{align} 
with $\alpha\in\{x,y,z\}$. Here $\mathds{1}$ denotes the identity matrix in spin space. The nonlinear terms $n$ and $\vec{F}$ are determined using $\bm{\Psi}$.
Here we focus our attention on a quasi-2D system with spatial coordinates $\vec{\rho}=(x,y)$, where $c_0$ and $c_1$ are the quasi-2D coupling constants.

Stationary solutions of the form $\Psi_{m}(\vec{\rho},t)=\psi_{m}(\vec{\rho}\,)e^{-i(\mu+m\lambda) t/\hbar}$ satisfy the time-independent Gross-Pitaevskii equation 
\begin{align}
(\mu\mathds{1}+\lambda \check{f}_{z})\bm{\psi}=\check{\mathcal{L}}\bm{\psi},\label{tiGPE}
\end{align}
 with nonlinear terms in $\check{\mathcal{L}}$ now evaluated with $\bm{\psi}$. Here $\mu$ and $\lambda$ are the chemical and magnetic potentials introduced as a Lagrange multipliers to conserve norm $N$ and $z$-magnetization $M_{z}$, respectively\footnote{These are defined by the   thermodynamic relations  $\mu=\left(\frac{\partial E}{\partial N}\right)_{\!S,V}$, and $\lambda=\left(\frac{\partial E}{\partial M_z}\right)_{\!S,V}$.  For the bulk EPP phase at $T=0$ (with entropy $S=0$)   $E=N(q+\frac{1}{2}c_{0}n_{b})+\frac{1}{2}c_{1}F_{z,b}M_{z}$ [from (\ref{HQZ}) and (\ref{spinH})], with $N=Vn_b$ and $M_z=VF_{z,b}$ for the appropriately dimensioned volume $V$.}. For a uniform EPP phase condensate of bulk density $n_b$ and magnetization density $F_{z,b}$ we have 
 \begin{align}\mu=c_0n_b+q,\label{mu}
 \end{align} and $\lambda=c_1F_{z,b}$  (e.g.~see \cite{Kawaguchi2012R}).
 
\subsection{Radial Spin-1 GPE}
An EPP phase vortex stationary state takes the form  [generalizing Eq.~(\ref{psiv})]:
\begin{align}
 \bm{\psi}_{\text{V}}(\rho,\phi) =e^{i\sigma_{\text{M}}\phi}\begin{bmatrix}
e^{-i\sigma_{\text{S}}\phi}\chi_1(\rho) \\
\chi_0(\rho) \\
e^{i\sigma_{\text{S}}\phi}\chi_{-1}(\rho) 
\end{bmatrix}=\check{C}(\phi)\bm{\chi}(\rho),\label{psiNSV}
\end{align}
where we have used the radial coordinate $\rho=\sqrt{x^{2}+y^{2}}$.   
 Here we take $\bm{\chi} =[\chi_1,\chi_0,\chi_{-1}]^{\mathrm{T}}$  as real, and have explicitly imposed the circulation on each component using $\check{C}=\text{diag} \{e^{i(\sigma_{\text{M}}-\sigma_{\text{S}})\phi},e^{i\sigma_{\text{M}}\phi},e^{i(\sigma_{\text{M}}+\sigma_{\text{S}})\phi}\}$.
  Substituting (\ref{psiNSV}) into Eq.~(\ref{tiGPE}),  gives a radial equation for the $\bm{\chi}$:
\begin{align}
\tilde{\mu}\bm{\chi} = \check{\mathcal{K}}\bm{\chi},\label{NSVGPE}
\end{align}
with
\begin{gather}
\check{\mathcal{K}}=T_{\rho}\mathds{1}+\tilde{p}\check{f}_{z}+\tilde{q}(\check{f}_{z}^{2}-\mathds{1})+c_{0}n\mathds{1}+c_{1}\sum_{\alpha}\left(\bm{\chi}^{\mathrm{T}}\check{f}_{\alpha}\bm{\chi}\right)\check{f}_{\alpha} \label{radialGPE}.
\end{gather} 
Here
\begin{align}
T_\rho &\equiv -\frac{\hbar^{2}}{2M}\frac{1}{\rho}\dv{\rho}\left(\rho\dv{\rho}\right),\\
\tilde{\mu}(\rho) &\equiv \mu_{b}-\frac{\hbar^{2}(\sigma_{\text{M}}^{2}+\sigma_{\text{S}}^{2})}{2M\rho^{2}},\label{mutilde}\\
\tilde{p}(\rho) &\equiv -\lambda-\frac{\hbar^{2}\sigma_{\text{M}}\sigma_{\text{S}}}{M\rho^{2}},\\
\tilde{q}(\rho) &\equiv q+\frac{\hbar^{2}\sigma_{\text{S}}^{2}}{2M\rho^{2}},
\end{align}
are the radial part of the kinetic energy operator, 
the effective chemical potential, 
and the effective linear and quadratic Zeeman shifts, respectively.
We have also subtracted $q$ off the single particle energy [see Eq.~(\ref{radialGPE})] for convenience, so that the adjusted chemical potential appearing in Eq.~(\ref{mutilde}) is given by
\begin{align}
\mu_b\equiv\mu-q=c_0n_b,\label{mub_fixed}
\end{align}
and is independent of $q$ [cf.~Eq.(\ref{mu})]. Here we take $\mu_b=c_0n_b$ as a useful characteristic energy scale of the system, with associated  healing length
\begin{align}
\xi_b=\frac{\hbar}{\sqrt{M\mu_b}}.
\end{align}

\begin{figure}[htbp] 
   \centering
   \includegraphics[width=3.4in]{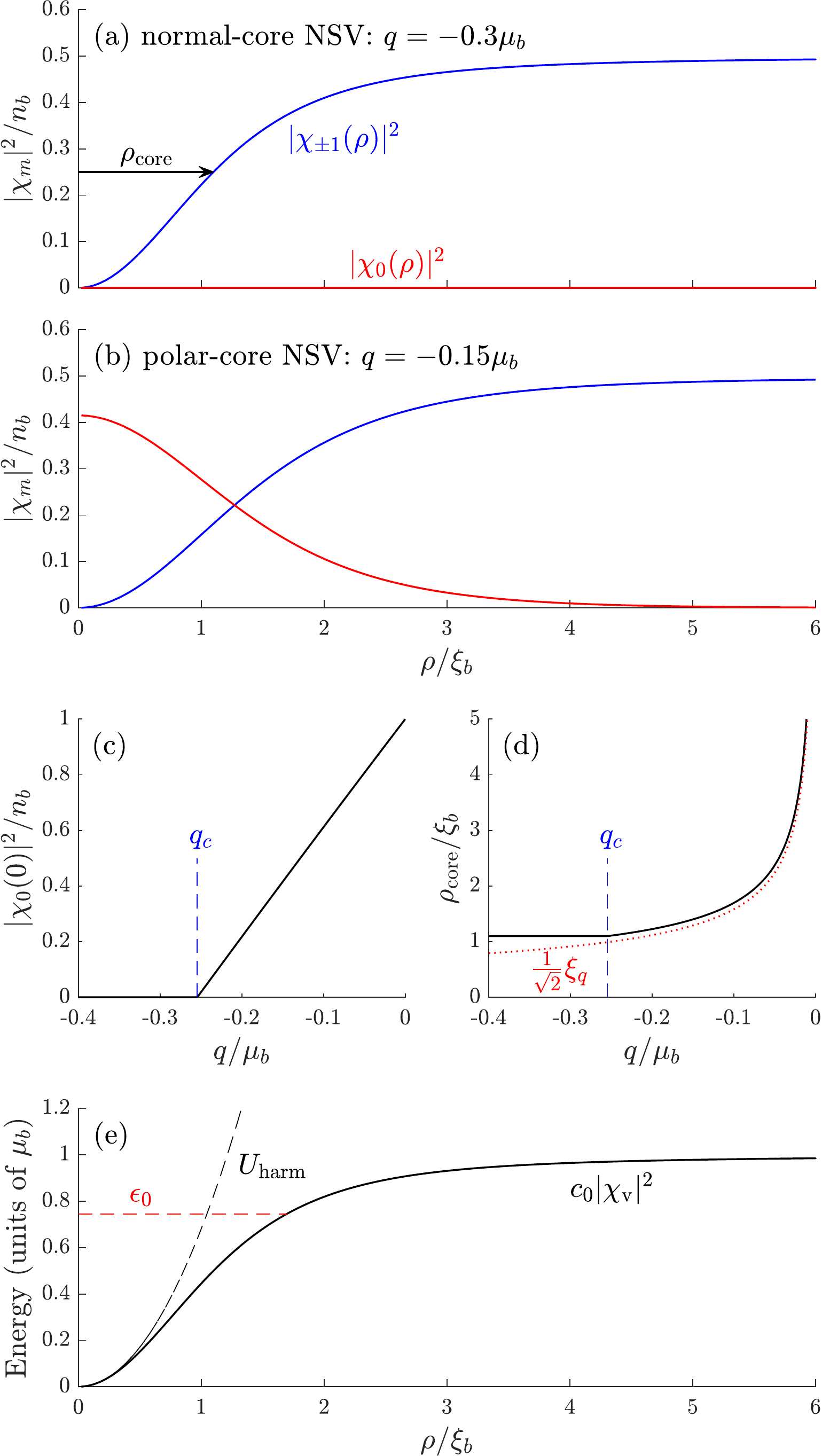} 
   \caption{ Stationary state component densities $|\chi_m(\rho)|^2$  of a NSV with (a) a normal core  and (b) a polar core. 
   (c) The central (peak) density of the $m=0$ component. (d) The core radius $\rho_{\text{core}}$ defined as the radius where $|\chi_{\pm 1}(\rho_{\text{core}})|^2=\frac{1}{4}n_b$ [see arrow in subplot(a)]. The red dotted line shows $\frac{1}{\sqrt{2}}\xi_q$.    (e) The effective potential $c_0|\chi_\mathrm{v}|^2$ for the core localized state $\psi_\mathrm{core}$ used to identify the critical point (see text). }
   \label{fig:vortprop}
\end{figure}

\subsection{Phases of a NSV}

In this work we consider a NSV with $(\sigma_{\text{M}},\sigma_{\text{S}})=(0,1)$. From our choice of $\mu_b$ being fixed [Eq.~(\ref{mub_fixed})], far away from the vortex core the system will approach the bulk value for number density , i.e.~$n_b$. Additionally,  we focus on the case $\lambda=0$ (and thus $\tilde{p}=0$), for which the bulk spin density is $F_{z,b}=0$.   In this case the NSV is completely unmagnetized\footnote{The discrete symmetry associated with the operation $\check{f}_z\to -\check{f}_z$ is explicitly broken for the ``biased'' case with $\lambda\neq 0$, where a magnetization ($F_z\neq 0$) is generally preferred. In this work, we consider the ``unbiased'' case with $\lambda=0$. Even for $\lambda=0$, a spontaneous magnetization can occur locally, e.g.~$F_z\neq 0$ in the core of a HQV. In this case, the spinor field is no longer represented by the real vector $\vec{d}$ like Eq. (\ref{psiPolar}).} with $\chi_1=-\chi_{-1}$, and the boundary conditions on the $\chi_m$ are:
\begin{align}
\chi_0^\prime(0)=0, \qquad&\chi_{0}(\rho\to\infty)=0,\label{chiBCs1}\\
\chi_{\pm 1}(0)=0, \qquad&\chi_{\pm 1}(\rho\to\infty)=\mp\sqrt{\frac{n_b}{2}},\label{chiBCs2}
\end{align}
where the prime denotes a derivative with respect to $\rho$.
We solve for the  stationary state solutions numerically using a gradient flow technique (e.g.~see \cite{Bao2013a}) with a finite difference implementation of the derivative operators and boundary conditions. We use an equally spaced radial grid of $N_\rho$ points $\rho_j=(j-\frac{1}{2})\Delta \rho$ with $1\le j\le N_\rho$.
We choose the point spacing $\Delta \rho$ to be much smaller than $\xi_b$ and typically use a maximum radius of $\rho_\mathrm{max}\approx410\xi_b$, with $N_\rho=8192$. We choose to implement the outer boundary conditions in Eqs.~(\ref{chiBCs1}) and  (\ref{chiBCs2}) as $\chi^\prime_{m}(\rho_\mathrm{max})=0$.

Because the NSV stationary solutions are  completely unmagnetized they are independent of the strength of the spin-dependent interaction. However, they depend on the quadratic Zeeman energy, and we find that there is a critical value 
\begin{align}
q_c\approx-0.2545\mu_b,\label{qc}
\end{align}
 [or $q_{c}\approx -0.3414\mu$, see Eq.~(\ref{mub_fixed})], which separates the normal-core and polar-core forms of the vortex.

\subsubsection{Normal-core NSV}\label{Sec:NCV}
For $q<q_c$  the stationary state  has $\chi_0=0$ and is otherwise is independent of $q$ [e.g., see Fig.~\ref{fig:vortprop}(a)]. In this regime the vortex profile is  $\chi_{-1}(\rho)= \tfrac{1}{\sqrt{2}}\chi_\text{v}(\rho) $, where   $\chi_\text{v}$ is the radial profile\footnote{Defining the scalar vortex state as $\psi_\text{v}(\vec{\rho}\,)=e^{i\phi}\chi_\text{v}(\rho)$.} of a single component (scalar) vortex in uniform system satisfying  
\begin{align}
\mu_b\chi_\text{v}= \left(T_{\rho}+\frac{\hbar^2}{2M\rho^2}+c_0|\chi_\text{v}|^2\right)\chi_\text{v},
\end{align}
with chemical potential $\mu_b$ used to ensure that   $|\chi_\text{v}|^2$ goes to $n_b$ as $\rho\to\infty$.

\subsubsection{Polar-core NSV}\label{Sec:PCV}
When $q>q_c$ the $\chi_0$ component is non-zero, and constitutes a polar core of the vortex  [e.g., see Fig.~\ref{fig:vortprop}(b)].
As $q$ further increases the density of the $\chi_0$ component and the core radius of the NSV  both increase, with $|\chi_0(0)|^2\to n_b$ and the core radius diverging as $q\to0$ [see Fig.~\ref{fig:vortprop}(c) and (d)].

We can qualitatively understand this behaviour by noting that the effective quadratic Zeeman energy for the system $\tilde{q}(\rho)$ is spatially dependent due to the vortex kinetic energy. 
Within a local density approximation the sign of $\tilde{q}$ determines local nematic spin order: the EPP phase occurs where $\tilde{q}<0$ and EAP phase occurs where $\tilde{q}>0$ (see Sec.~\ref{Sec:GenFormalism}). We see that the effect of the vortex kinetic energy is to transition a central region of radius $\rho_{\text{core}}\sim\xi_q=\hbar/\sqrt{2M|q|}$ into the EAP phase, where  $\xi_q$ is defined  by  $\tilde{q}(\xi_q)=0$.  In Fig.~\ref{fig:vortprop}(d) we observe that $\xi_q$ provides a good estimate of the core size for $q>q_c$, noting that for $q<q_c$ the local density approximation breaks because of finite-size effects in the vortex core.

\subsubsection{Identifying the critical point}\label{Seccritpoint}
Near the critical point $\chi_0$ is small (i.e.~$|\chi_0|^2\ll n_b$) and a single-particle treatment of this component can be employed. In this regime the  $m=0$ component of the GPE (\ref{NSVGPE}) reduces to (neglecting the nonlinear terms in $\chi_0$):
\begin{align}
\mu_b\chi_0=(T_{\rho}+c_0|\chi_\text{v}|^2-q)\chi_0,\qquad |\chi_0|^2\ll n_b,\label{GPE0lin}
\end{align} 
where we have set $n=|\chi_\text{v}|^2$ (i.e.~the normal-core NSV density, see Sec.~\ref{Sec:NCV}) in the interaction term. 
The lowest energy eigenstate of the linear operator  $T_{\rho}+c_0|\chi_\text{v}|^2-q$  is a core localized state of energy\footnote{\label{FNcoremode}The lowest energy eigenstate of $T_\rho+c_0|\chi_\text{v}|^2$ is  $\psi_\text{core}$ with a numerically determined energy of $\epsilon_0=0.7455\mu_b$. This state is bound within the vortex core [see Fig.~\ref{fig:vortprop}(e)].   Note that  the harmonic approximation to $c_0|\chi_\text{v}|^2$, i.e.~$U_\text{harm}=\mu_b(\Lambda\rho/\xi_b)^2$, with $\Lambda= 0.8249$ \cite{Bradley2012a}, is inaccurate and  predicts  $\epsilon_0=1.14\mu_b$, which exceeds the core well-depth of $\mu_b$.}
$\epsilon_\text{core}=0.7455\mu_b-q$. The system will ``condense" into this state when the chemical potential,  $\mu_b$, exceeds $\epsilon_\text{core}$. Thus we take $\mu_b=\epsilon_\text{core}$ to define the critical value of the quadratic Zeeman energy, yielding a value of $q_c$ in agreement with the value identified from the GPE calculations for the NSV [Eq.~(\ref{qc})].

\section{Linear stability of the NSV}\label{Sec:Excitations}
We now examine the excitations of the NSV by directly solving the BdG equations. We begin by deriving the BdG equations for an EPP phase vortex. We then use numerical solutions of these equations to quantify NSV stability.

\subsection{Bogoliubov-de Gennes equations} 
To derive the BdG equations we introduce a time-dependent fluctuation $\delta\bm{\psi}(\vec{\rho},t)=\bm{\psi}(\vec{\rho},t)-\bm{\psi}_{\text{V}}(\vec{\rho}\,)$ on the vortex stationary state $\bm{\psi}_{\text{V}}$ [see Eq.~(\ref{psiNSV})]:
\begin{align}
\delta\bm{\psi}(\vec{\rho},t)=\check{C}(\phi)\sum_{\nu,\eta}\left[\beta_{\nu,\eta}\bm{u}_{\nu,\eta}-\beta_{\nu,\eta}^*\bm{v}_{\nu,\eta}^{*}\right],\label{dpsi_BdG}
\end{align}
where $\beta_{\nu,\eta}$ are arbitrary (small) linearization amplitudes, and
\begin{align}
\bm{u}_{\nu,\eta}(\vec{\rho},t)=e^{-iE_{\nu,\eta}t/\hbar+i\eta\phi}\tilde{\bm{u}}_{\nu,\eta}(\rho),\\
\bm{v}_{\nu,\eta}(\vec{\rho},t)=e^{-iE_{\nu,\eta}t/\hbar+i\eta\phi}\tilde{\bm{v}}_{\nu,\eta}(\rho).\label{vform}
\end{align}
Here we have introduced the radial quasiparticle amplitudes  $\{\tilde{\bm{u}}_{\nu,\eta}(\rho),\tilde{\bm{v}}_{\nu,\eta}(\rho)\}$ and eigenvalues $E_{\nu,\eta}$, with $\eta$ being the quantum number  associated with the $z$-component of angular momentum (relative to the condensate) and $\nu$ representing the remaining quantum numbers. Linearizing the time-dependent GPE (\ref{tdGPE}) we obtain that the quasiparticle amplitudes satisfy the Bogoliubov-de Gennes (BdG) equation
\begin{align}
\begin{bmatrix}
\check{\mathcal{K}}^+_\eta +\check{X}_1 & -\check{X}_2 \\
\check{X}_2^* & -(\check{\mathcal{K}}^-_\eta+\check{X}_1)^*
\end{bmatrix}
\begin{bmatrix}
\tilde{\bm{u}}_{\nu,\eta}\\
\tilde{\bm{v}}_{\nu,\eta}
\end{bmatrix}
=E_{\nu,\eta}
\begin{bmatrix}
\tilde{\bm{u}}_{\nu,\eta}\\
\tilde{\bm{v}}_{\nu,\eta}
\end{bmatrix},
\end{align}
where nonlinear terms in
\begin{align}
\check{\mathcal{K}}^\pm_\eta&=\check{\mathcal{K}}+\frac{\hbar^{2}}{2M\rho^{2}}\left[\left(\eta^{2}\pm 2\sigma_{\text{M}}\eta\right)\!\mathds{1}\mp 2\sigma_{\text{S}}\eta \check{f}_{z}\right]-\tilde{\mu}\mathds{1},
\end{align}
are evaluated with $\bm{\psi}_{\text{V}}$, and 
\begin{align}
\check{X}_1&=c_0\bm{\chi}\bm{\chi}^{\mathrm{T}}+c_1\sum_{\alpha}\check{f}_{\alpha}\bm{\chi}\bm{\chi}^{\mathrm{T}}\check{f}_{\alpha}, \\
\check{X}_2&=c_0\bm{\chi}\bm{\chi}^{\mathrm{T}}+c_1\sum_{\alpha}\check{f}_{\alpha}\bm{\chi}\bm{\chi}^{\mathrm{T}}\check{f}_{\alpha}^{*}.
\end{align}

Because of the symmetry of the radial BdG equation for a solution (i) $(E,\eta,\tilde{\mathbf{u}},\tilde{\mathbf{v}})$ there are potentially three additional solutions which relate to the first as: (ii) $(-E,-\eta,\tilde{\mathbf{v}},\tilde{\mathbf{u}})$, (iii) $(E^*,\eta,\tilde{\mathbf{u}}^*,\tilde{\mathbf{v}}^*)$, and 
(iv) $(-E^*,-\eta,\tilde{\mathbf{v}}^*,\tilde{\mathbf{u}}^*)$ . If the eigenvalue is real, then $\tilde{\mathbf{u}}$ and $\tilde{\mathbf{v}} $ can also be taken real (see Ref.~\cite{Takahashi2009a})  and only (i) and (ii) are unique solutions. Furthermore quasiparticle amplitudes with real non-zero eigenvalues can be normalized to $\pm1$ as
\begin{align}
\int_0^\infty 2\pi \rho\,\dd{\rho}\,(\bm{u}_{\nu,\eta}^\dagger\bm{u}_{\nu',\eta'} -\bm{v}_{\nu,\eta}^\dagger\bm{v}_{\nu',\eta'})=\pm\delta_{\nu\nu'}\delta_{\eta,\eta'}. \label{QPnorm}
\end{align}
In the description of equilibrium condensates only positively normalized quasiparticles are considered to be physical. We note that the partner (ii) to a positively normed quasiparticle (i) has negative norm.  

Here we are particularly interested in quasiparticles with complex eigenvalues where all the symmetries (i)-(iv) furnish unique solutions. If  $\text{Im}(E)>0$ then the solution (i) is dynamically unstable (exponentially growing in time), and so is the partner solution (iv), while (ii) and (iii) are exponentially decaying solutions. Examining the effect of the unstable mode perturbation $\delta \bm{\psi}$ (\ref{dpsi_BdG}) we see modes (i) and (iv) are identical perturbations, so here we can choose to focus on solution (i).

\begin{figure}[htbp] 
   \centering
   \includegraphics[width=3.4in]{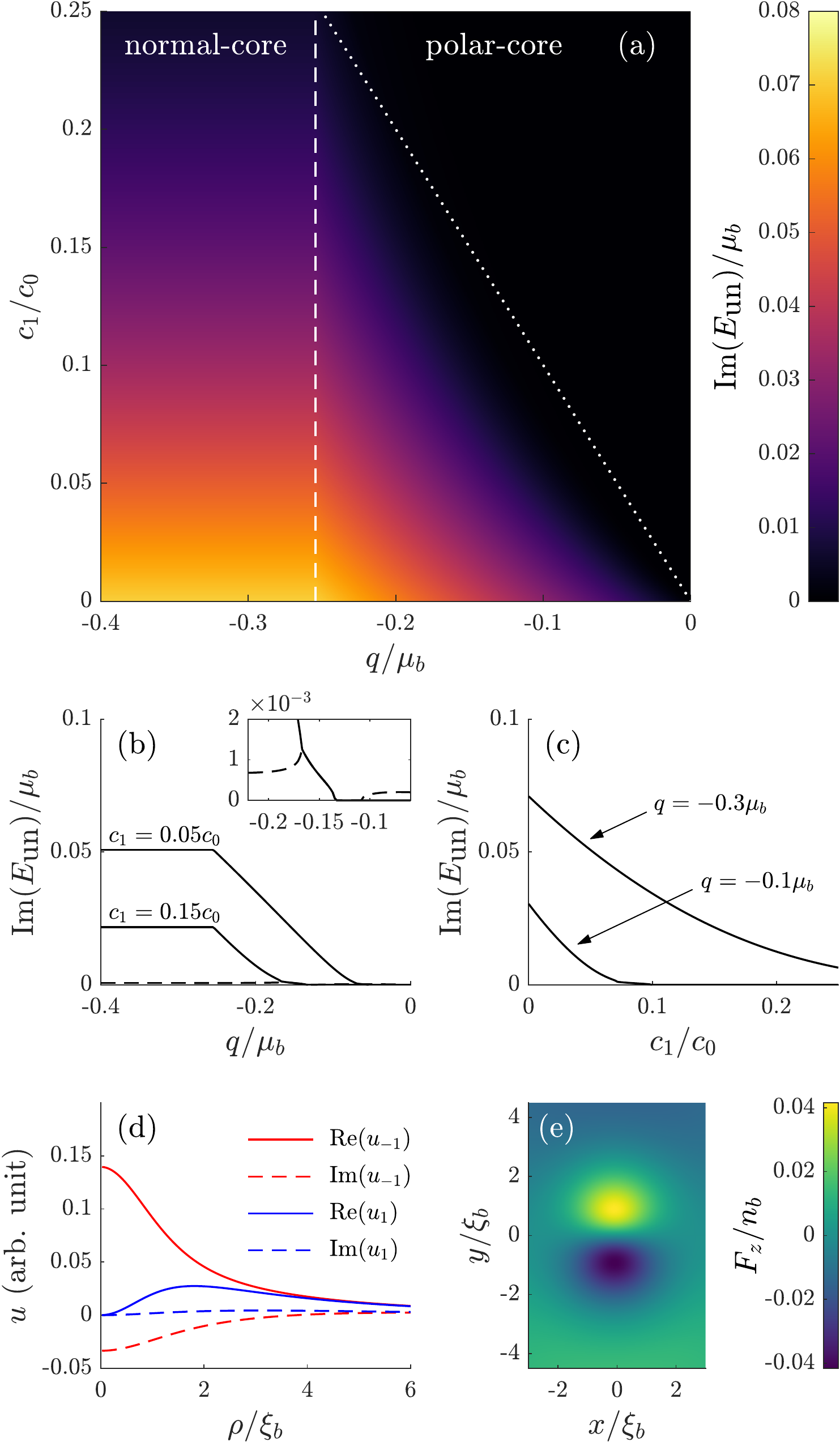}  
   \caption{Unstable (splitting) mode of the NSV. (a) Phase diagram showing the imaginary part of the eigenvalue of the unstable mode with $\eta=\pm1$. Dotted line indicates the simple model for instability boundary based on counter-superflow instability.  
Several slices through the phase diagram showing the strength of instability as (b) $q$ varies, and (c) as $c_1$ varies.  In (b) the dashed line shows the zero mode for $c_1=0.15c_0$. The inset shows the behaviour of the $c_1=0.15c_0$ modes near where the instability vanishes. (d) The non-zero $u$-components of the $\eta=-1$ unstable mode for $q=-0.3\mu_b$ and $c_1=0.05c_0$. (e)  The $z$-component of the spin density of the NSV after the unstable mode shown in (d) is added to the stationary vortex state, revealing that the vortices of the two components spatially separate. }
   \label{fig:Excites}
\end{figure}
 
\subsection{Dynamic instability phase diagram}

The BdG results can reveal two types of instabilities for the NSV: (i) A dynamic instability revealed by a solution with a complex eigenvalue  indicating that the respective eigenmode will exponentially grow with time. (ii) A Landau instability marked by a (positively normed) solution with a negative real eigenvalue, such that the system could reduce its energy if some dissipative mechanism allowed transfer of population into this state.

We have numerically calculated the BdG spectrum of the NSV over a wide parameter regime. We find that for $q<0$ the NSV only exhibits dynamic instabilities which occur in excitations with $\eta=\pm1$, arising from modes that are localized in the vortex core\footnote{The centrifugal term in the $m=-\eta$ ($m=\eta$) component of the operator $\check{\mathcal{K}}_\eta^{+}$ ($\check{\mathcal{K}}_\eta^{-}$) vanishes for $\eta=\pm1$,  allowing the corresponding component of the $\bm{u}_{\nu,\eta}$ ($\bm{v}_{\nu,\eta}$) amplitude to develop amplitude in the NSV core.} [see Fig.~\ref{fig:Excites}(d)]. The growth of these unstable modes causes the $m=\pm1$ cores of the NSV to separate [see Fig.~\ref{fig:Excites}(e)], thus initiating the dissociation of the NSV into two HQVs (see Sec.~\ref{Sec:IntroNSV}).  

In Fig.~\ref{fig:Excites}(a) we present a stability phase diagram quantifying the dynamic instability of the NSV (i.e.~showing the imaginary part of the eigenvalue of the dynamically unstable mode) as $q$ and $c_1$ vary. In general the imaginary part is always relatively  small ($\lesssim10^{-1}\mu_b$), so we expect the instability to manifest slowly in the system dynamics (see Sec.~\ref{Sec:dynamics}).
 The dynamic instability is seen to depend on both $q$ and $c_1$. It is independent of $q$ for the normal-core NSV (i.e.~$q<q_c$), but reduces with increasing $q$  in the polar-core regime [Fig.~\ref{fig:Excites}(b)].  The instability also decreases with increasing $c_1$  [Fig.~\ref{fig:Excites}(c)].

\subsubsection{$q$-dependence of instability: Pinning effect of polar core}
For the polar-core NSV the magnitude of the dynamic instability decreases with increasing $q$. We interpret this as a \textit{pinning} effect of the polar core that helps bind the two component vortices together, and thus stabilize the NSV. The effects of pinning (e.g.~due to an external potential, the other superfluid component, or thermal component) has previously been considered as a mechanism for stabilizing vortices  (e.g.~see \cite{Isoshima1999a,Isoshima2002a,Simula2002a,hayashi2013instability}). 

To quantify the pinning effect we consider a polar-core NSV solution $\bm{\chi}_\text{PC}=[-\chi_{-1},\chi_0,\chi_{-1}]^\mathrm{T}$. From this we can project out the $m=0$ component to arrive at an effective normal-core NSV $\bm{\chi}_{\text{ENC}}=[-\chi_{-1},0,\chi_{-1}]^\mathrm{T}$ that is a stationary solution for the same $q$ value if we add the scalar potential $U_\text{pin}=c_0|\chi_0|^2$ to the GPE\footnote{I.e.~add a term $U_\text{pin}\mathds{1}$ to $\mathcal{L}_\rho$ to compensate for the reduction of $c_0n$ by the removal of the $\chi_0$ component.}. The unstable modes in the BdG analysis of $\bm{\chi}_\text{ENC}$ (including the pinning potential) have larger imaginary parts than those obtained for $\bm{\chi}_\text{PC}$. This demonstrates that there is an intrinsic spin-dependent aspect to the pinning stabilization. If we artificially increase the strength of the pinning potential (i.e.~set $U_\text{pin}\to\gamma U_\text{pin}$, with $\gamma>1$) then eventually the dynamical instability is suppressed.

\subsubsection{$c_1$-dependence of instability: Counter-superflow instability}\label{Sec:CSI}
Counter-superflow instability involves the breakdown of spin-superfluidity when the relative velocity of two miscible superfluids exceeds a critical value \cite{Law2001a,Yukalov2004a,Takeuchi2010a,Suzukia2010a,Abad2015a,Hoefer2011a,Hamner2011a,Fujimoto2012a,Zhu2015a,Kim2017a}, and affords a qualitative understanding of the dependence of the NSV instability on $c_1$.
For a uniform spinor condensate in the EPP phase the critical relative velocity for the onset of the instability is $v_\text{crit}=2\sqrt{c_1n/M}$  \cite{Zhu2015a}. We can apply this criteria to NSV using the local density approximation (similar to the treatment presented in Ref.~\cite{Ishino2013a}).
 The relative velocity arises from the counter-rotating vortices in the $m=\pm1$ components and varies radially as $v_\text{rel}(\rho)=\frac{2\hbar}{M\rho}$.  
Approximating the NSV density by the background value $n_b$ we identify the critical radius  
 $\rho_\text{crit}=\sqrt{\frac{c_0}{c_1}}\xi_b$,
from the condition $v_\text{rel}(\rho_\text{crit})=v_\text{crit}$. For $\rho<\rho_\text{crit}$ the relative velocity exceeds $v_\text{crit}$ and counter-superflow instability is activated. This analysis suggests that the instability will be stronger closer to the core, consistent with unstable modes being localized near the core [see Fig.~\ref{fig:Excites}(d)].  Also, as $c_1/c_0$ and hence the critical velocity increases, $\rho_\text{crit}$ decreases, suggesting that the instability should be weaker.  

This analysis does not apply to the core region  as here the density varies rapidly so that the local density approximation is inapplicable.   If we assume that the instability is suppressed once the critical radius is comparable to the vortex core size we can quantify a stability boundary for the system.  In the polar-core regime we estimate the core size as $\xi_q$, and the critical radius is equal to this when 
\begin{align}
\frac{c_1}{c_0}=-\frac{q}{\mu_b}=-\frac{q}{\mu-q},\qquad\mbox{(stability boundary)}. 
\end{align}
 This is shown as a dotted line in Fig.~\ref{fig:Excites}(a), and is seen to reasonably characterize the boundary of instability in the polar-core regime.

  \begin{figure}[htbp] 
   \centering
   \includegraphics[width=3.6in]{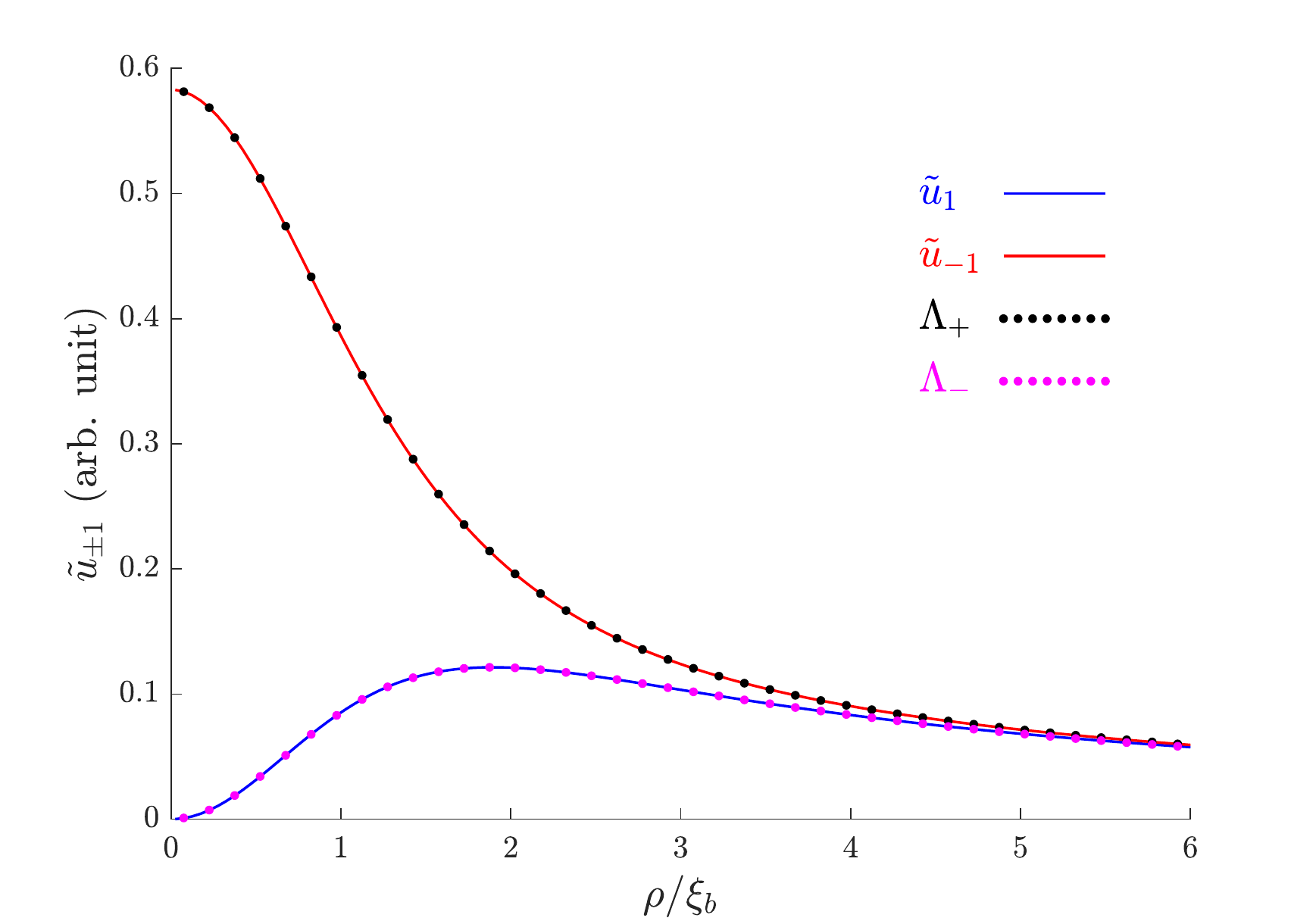}  
   \caption{The non-trivial $u$-components of a $\eta=-1$ zero energy mode and the functions 
   $\Lambda_\pm$. The same scaling factor is applied to the $u$-amplitudes. Note the nontrivial  $v$-components for the zero energy mode are given by $\tilde{v}_\pm=\tilde{u}_\mp$. Parameters are as in Fig.~\ref{fig:Excites}(d).  }
   \label{fig:zeromode}
\end{figure}

\subsection{Zero energy modes from broken translational symmetry}  
For $\eta=\pm1$ we find that in addition to the unstable modes there are also zero-energy modes. These modes often reveal a broken symmetry in the system. 
These new zero modes are in addition to the usual $\eta=0$ zero energy mode, which can be written in terms of the condensate mode as $\{\bm{u}_{0,0},\bm{v}_{0,0}\}=\{\bm{\chi} ,-\bm{\chi}\}$  and is associated with the breaking of the gauge symmetry. The new zero energy modes with $\eta=\pm1$ are associated with the breaking of translational symmetry in our uniform system by the presence of the NSV.  These modes are localized in the core, similar to the unstable modes, but have a flat phase profile [see Fig.~\ref{fig:zeromode} and cf.~Fig.~\ref{fig:Excites}(d)]. Whereas the unstable mode causes the $m=\pm1$ component vortices to separate, the zero energy mode causes both to translate together.
For a scalar condensate a similar zero mode emerges and in the case of a vortex line (i.e.~finite  $z$-extent), it is associated with the Kelvin-wave spectrum of helical modes that propagate along the vortex line. 

 We can develop an analytic expression for these zero energy modes that we compare to the numerical solution of the BdG equations. 
We restrict our attention to a normal-core  NSV $(\sigma_\text{M},\sigma_\text{S})=(0,1)$,  which has the form [see Eq.~(\ref{psiNSV})] $\bm{\psi}_V =[-\chi_{-1}(\rho)e^{-i\phi},0,\chi_{-1}(\rho)e^{i\phi}]^\mathrm{T}$. Considering the vortex to be displaced by a small amount $\text{d}\vec{\rho}$ such that the change in the condensate wavefunction is $\delta \bm{\psi}=\text{d}\vec{\rho}\cdot\vec{\nabla}\bm{\psi}_V$, we obtain for the change in the nontrivial $m=\pm1$ components
\begin{align}
\delta{\psi}_{\pm1}= e^{\mp i\phi}\left[-\left(\dd{x}-i\dd{y}\right)\Lambda_{\pm}e^{i\phi}+\left(\dd{x}+i\dd{y}\right)\Lambda_{\mp}e^{-i\phi}\right]
 \label{eq_dspi_z}
\end{align}
where 
\begin{align}
\Lambda_{\pm}(\rho)= \frac{1}{2}\left(\frac{\chi_{-1}}{\rho}\pm\frac{\text{d}\chi_{-1}}{\text{d}{\rho}}\right).\label{Lambda}
\end{align}
By inspecting the form of the BdG linearization [Eqs.~(\ref{dpsi_BdG})-(\ref{vform})] we see that the vortex shift can be mapped to quasiparticle amplitudes with $\eta=\pm1$. For definiteness we consider a zero energy mode of the BdG solution  with $\eta=-1$, which we denote as $\{\tilde{\bm{u}},\tilde{\bm{v}}\}$, with amplitude $\beta$, such that
\begin{align}\delta{\psi}_{\pm1}=e^{\mp i\phi}[ {\beta}\tilde{u}_{\pm1}e^{-i\phi}- {\beta}^*\tilde{v}_{\pm1}^*e^{i\phi}].
\end{align}
In comparison to Eq.~(\ref{eq_dspi_z}) we observe $\tilde{u}_{\pm1}\sim\Lambda_{\mp}$, and
$\tilde{v}_{\pm1}\sim\Lambda_{\pm}$, with the complex  amplitude $\beta$ determining the vortex displacement.  In Fig.~\ref{fig:zeromode} we show the numerically calculated BdG zero energy mode result and the functions $\Lambda_{\pm}$ (obtained from the vortex solution), showing they coincide.

\subsection{Numerical considerations}
It is challenging to numerically calculate the $\eta=\pm1$ unstable and zero modes accurately.  One issue is that our grids are of finite spatial extent $\rho_\mathrm{max}=410\xi_b$. This is problematic for the zero energy mode which decays rather slowly [noting that $\Lambda_{\pm}\sim\rho^{-1}$, see Eq.~(\ref{Lambda})]. Additionally, the 2D radial Laplacian is difficult to evaluate accurately with finite differences, particularly near $\rho=0$  \cite{Arsoski2015a,Laliena2018a}.  In the BdG equations for $\eta=\pm1$  the unstable and zero energy modes are particularly sensitive to these issues. We have found that second order finite difference schemes (including the improved schemes presented in  Refs.~\cite{Arsoski2015a,Laliena2018a}) require an  impractically large number of grid points to obtain accurate results. This motivated us to implement an 8th order finite difference scheme, which is the basis of the results we present. With this scheme small artefacts are still apparent such as the zero mode having a small imaginary part [e.g.~see Fig.~\ref{fig:Excites}(b) and inset], causing it to couple to the unstable mode. These artefacts reduce as the range and point density of the numerical grid increase.

\section{Dynamical simulations}\label{Sec:dynamics}

\begin{figure}[t]
	\centering
	\includegraphics[width=\linewidth]{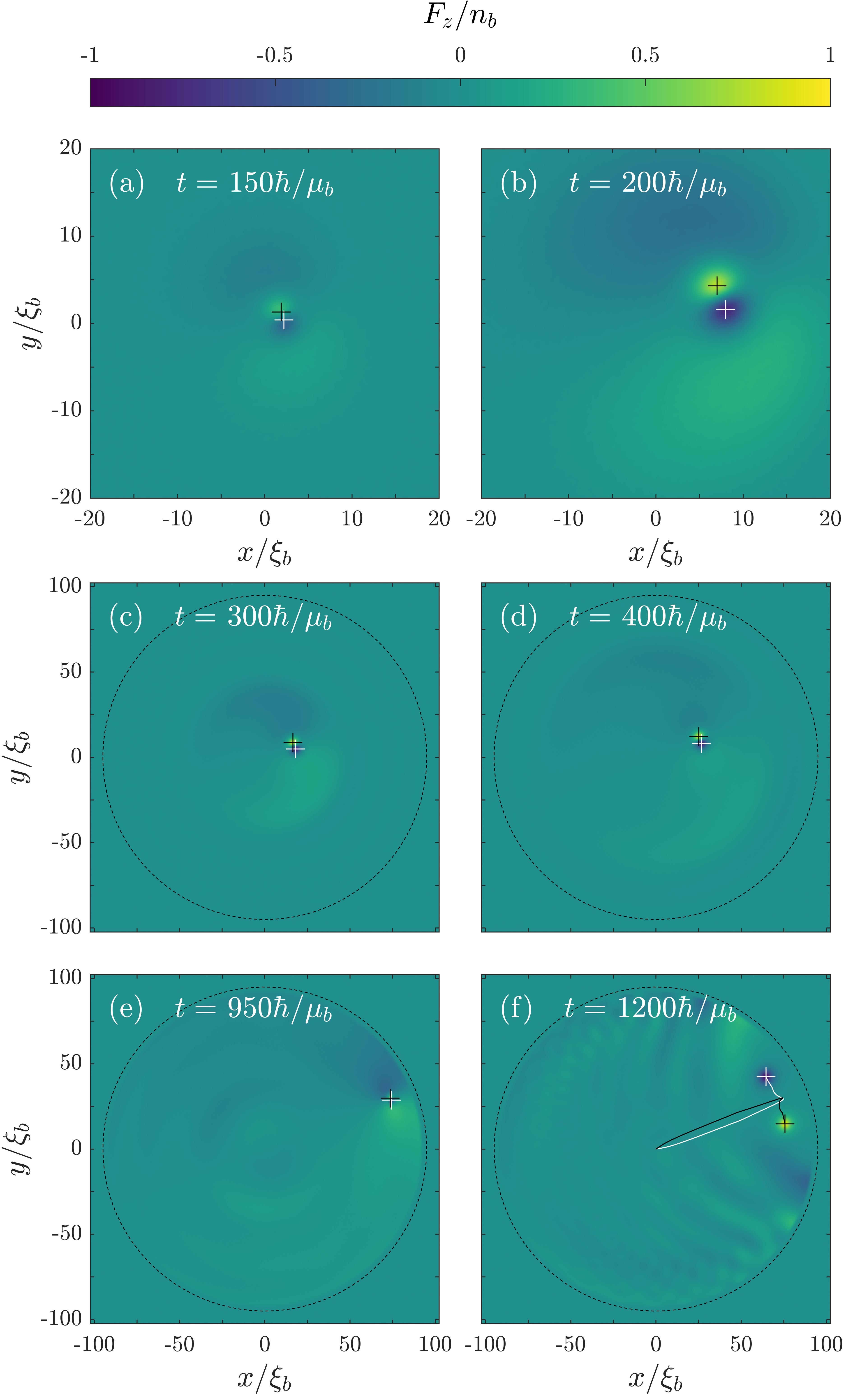}
	\caption{(a)-(f) Time-evolution of the $z$-component of the spin-density of a normal-core NSV in a circular flat-bottomed trap.
	The vortices in the $m=1$ (white cross) and $m=-1$ (black cross) components are indicated. Subplots (a) and (b) zoom in to reveal the vortex dynamics in the central region, whereas (c)-(f) show the full simulation domain and indicate the trap boundary (black dashed line). In (f) the vortex trajectories are also shown for the vortices in the $m=1$ (white line) and $m=-1$ (black line) components.
	Other simulation parameters are $q=-0.3\mu_{b}$,  $c_{1}=0.05c_{0}$, $U_0=100\mu_b$, and $R=95\xi_b$ is the radius.}
	\label{fig:NSVdynamics}
\end{figure}

We further investigate the stability of NSVs by simulating their dynamics using the time-dependent GPE (\ref{tdGPE}).
The simulations are performed with a flat-bottomed circular trap (i.e.~scalar potential added to the GPE) of the form
\begin{align}
U(\rho)=\frac{1}{2}U_0\left[\tanh\left(\frac{\rho-R}{\xi_b}\right)+1\right],
\end{align}
where $U_0$ is the trap depth, and $R$ is the radius. The initial state is a NSV centered at the origin, obtained by solving the radial GPE [Eq.~(\ref{NSVGPE}) including $U(\rho)$] in the flat-bottomed trap potential using the approach described in Sec.~\ref{Sec:NVS}. This state is interpolated onto a uniform $2048\times 2048$-point 2D grid with spacing $0.1\xi_{b}$. A small amount of complex Gaussian noise is added to seed any instabilities in the dynamics. This is first prepared as white noise (on the position space grid), then restricted in reciprocal space to have maximum wave-number $8\xi_{b}^{-1}$, and finally spatially filtered to the region within the trap. Typically adding this noise to the initial state causes a $0.005\%$ increase in the wavefunction norm and a $0.05\%$ increase in the system energy. We time-evolve the resulting state using the second-order symplectic method described in Ref.~\cite{Symes2016a}.
 
We use the $z$-spin density to illustrate the evolution of a normal-core NSV in Fig.~\ref{fig:NSVdynamics}.  Initially $F_z$ is zero (to the level of the noise) but as the component vortices separate, clear structure develops. The vortex core in the $m=1$ component is filled by the $m=-1$ component, and thus appears as a negative $F_z$ peak. Similarly the vortex core in the $m=-1$ component appears with a positive $F_z$ peak. As time progresses the component vortex separation tends to increase and they move away from trap centre. Eventually the vortices approach the boundary where they undergo a sudden change in their motion causing the appreciable emission of spinwaves  [see Fig.~\ref{fig:NSVdynamics}(f)]. In contrast for a polar-core NSV with $q>-0.05\mu_b$ and $c_1n_b>0.05\mu_b$ [which is stable according to the BdG analysis, see Fig.~\ref{fig:Excites}(a) and (b)] we observed the component vortices to remain together at the origin for the entire evolution (i.e.~up to $t_{\mathrm{final}}=1200\hbar/\mu_b$).

\begin{figure}[t]
	\centering
	\includegraphics[width=\linewidth]{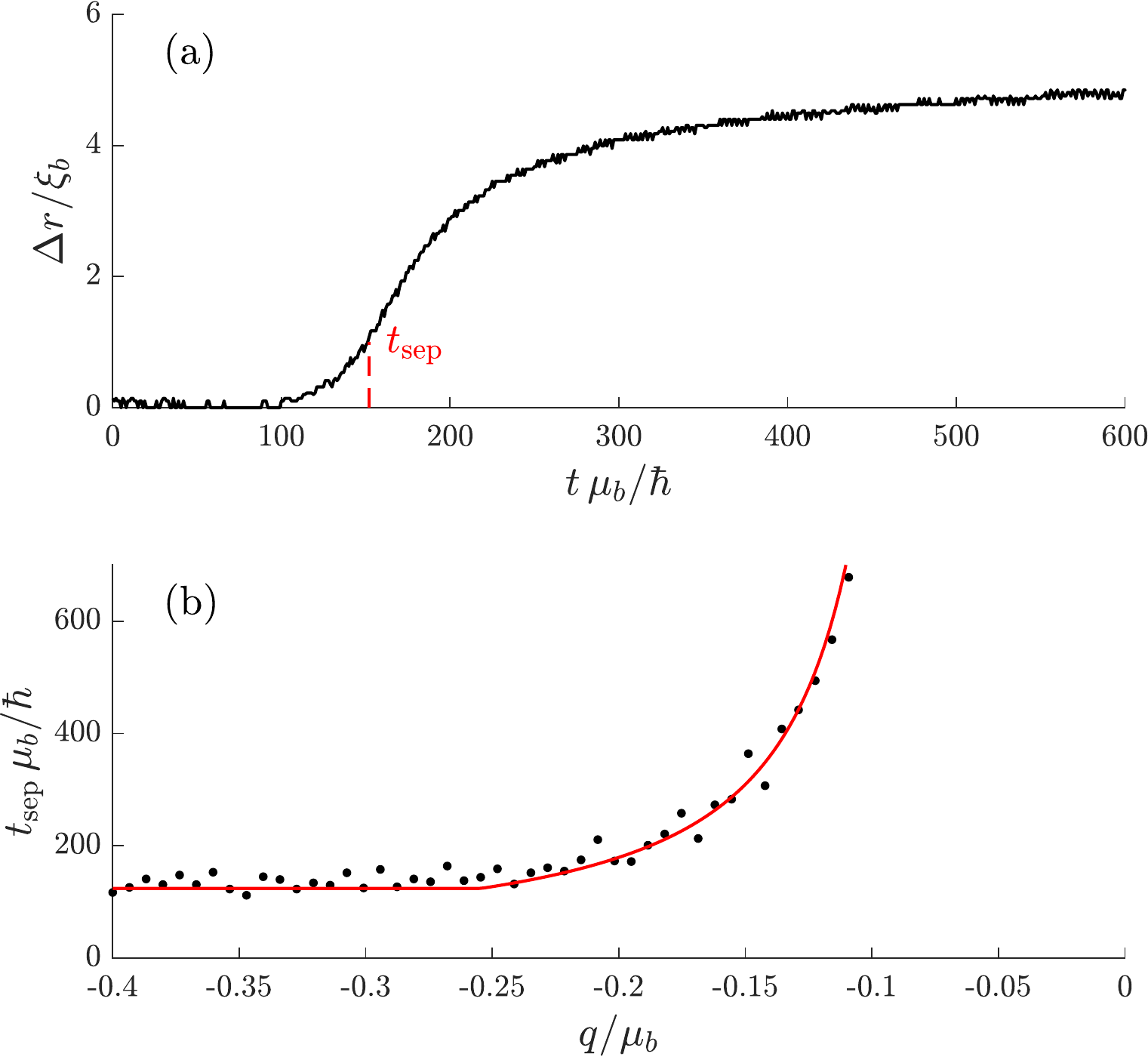}
	\caption{  (a) Vortex separation $\Delta r$ of a normal-core NSV, obtained with parameters $q=-0.3\mu_{b}$, $c_{1}=0.05c_{0}$. (b) Quadratic Zeeman dependence of the separation time $t_{\text{sep}}$ for $c_{1}=0.05c_{0}$. Black dots indicate numerical results. The red line is a fit to the BdG results, given by $4.07/E_{\text{un}}$, where $E_{\text{un}}$  is shown in Fig.~\ref{fig:Excites}(b).}
	\label{fig:NSVdynamics2}
\end{figure}

The above results motivate us to quantify the instability of the NSV in terms of the rate that the component vortices separate. 
In Fig.~\ref{fig:NSVdynamics2}(a) we show the evolution of the distance between the component vortices ($\Delta r$) for the case examined in Fig.~\ref{fig:NSVdynamics} over the initial period of its dynamics (i.e.~well before the boundary collision occurs).
We identify the separation time $t_{\text{sep}}$ as the time when $\Delta r$ first exceeds  {$\xi_{b}$}. In Fig.~\ref{fig:NSVdynamics2}(b) we show  $t_{\text{sep}}$ obtained from simulations conducted over a range of $q$ values. Here we see that $t_{\text{sep}}$ increases with $q$ for $q>q_c$, and appears to diverge  as $q$ approaches $-0.1\mu_b$. These results are consistent with the BdG analysis [see Fig.~\ref{fig:Excites}] if we identify $t_{\text{sep}}$ as scaling with $\hbar/|E_{\text{un}}|$, where $E_{\text{un}}$ is the (imaginary) eigenvalue of the dynamically unstable mode. A comparison to the BdG results is presented in Fig.~\ref{fig:NSVdynamics2}(b) and is seen to have good quantitative agreement.

\section{Discussion and conclusions}\label{Sec:conclude} 
Here we have presented a description of the NSV, outlining the stationary state properties  and a transition between a normal-core and polar-core form occurring at a critical value of the quadratic Zeeman energy. The NSV generally is unstable to dissociating into two HQVs. Using  a BdG analysis we quantify this instability, and find that it can be reduced by increasing the strength of the spin-dependent interactions. For the polar-core NSV the instability also decreases by increasing the quadratic Zeeman energy.

It should be possible to controllably produce NSVs using established experimental schemes involving magnetic and optical fields \cite{Shin2004a,Chen2018a}, which would allow the properties of individual NSVs to be studied. It is also interesting to ask if NSVs could play a role in the non-equilibrium dynamics of an antiferromagnetic spinor condensate quenched into the EPP phase. To date studies have considered the role of HQVs  (e.g.~\cite{Kang2017a,Symes2017b}), however we have identified regimes where NSVs are stable (or quasi-stable) defects. In such regimes they may be important in the description of phenomena such as phase ordering and quantum turbulence. 
We note that the easy-plane ferromagnetic spinor condensate similarly supports two types of vortices, with the dominant vortex type determining the universal ordering dynamics \cite{Kudo2015a,Williamson2016a,Schmied2019a}.

\begin{acknowledgments} 
APCU, DB, and PBB acknowledge support from the Marsden Fund of the Royal Society of New Zealand. HT was supported by JSPS KAKENHI Grant Numbers JP17K05549,
JP18KK0391, JP20H01842, and in part by the OCU ``Think globally, act
locally'' Research Grant for Young Scientists 2019 through the hometown
donation fund of Osaka City.
\end{acknowledgments}

% \bibliography{spinbib}


\end{document}